\documentclass[10pt,conference]{IEEEtran}

\usepackage{cite}
\usepackage{hyperref}
\usepackage{booktabs}
\usepackage{graphicx}
\usepackage{array}
\usepackage{microtype}
\usepackage{url}
\usepackage[most]{tcolorbox}
\usepackage{enumitem}
\usepackage{multirow}

\newtcolorbox{rqanswerbox}{
  enhanced,
  breakable,
  pad at break*=1mm,
  colback=gray!14,
  colframe=gray!65,
  boxrule=0.8pt,
  arc=2mm,
  left=6pt,
  right=6pt,
  top=6pt,
  bottom=6pt,
  before skip=8pt,
  after skip=8pt
}

\newcommand{\Description}[1]{}

\newcommand{\tool}{SDA}

\begin{document}

\title{Skills Are Not Islands: Measuring Dependency and Risk in Agent Skill Supply Chains}

\bstctlcite{IEEEexample:BSTcontrol}


\author{
\IEEEauthorblockN{Changguo Jia\IEEEauthorrefmark{1}, Tianqi Zhao\IEEEauthorrefmark{2},
Runzhi He\IEEEauthorrefmark{1}, and Minghui Zhou\IEEEauthorrefmark{1}}
\IEEEauthorblockA{\IEEEauthorrefmark{1}Peking University, Beijing, China\quad
\IEEEauthorrefmark{2}Zhongguancun Laboratory, Beijing, China}
\IEEEauthorblockA{jiachangguo@stu.pku.edu.cn, 
zhaotq@zgclab.edu.cn, rzhe@pku.edu.cn, zhmh@pku.edu.cn}
\IEEEauthorblockA{Corresponding author: Minghui Zhou (zhmh@pku.edu.cn)}
}

\maketitle

\begin{abstract}
Agent skills package reusable operational knowledge for Large Language Model (LLM) agents-but as they grow in scope, they become dependency-bearing artifacts whose identities, versions, and provenance remain largely implicit. This opacity is not merely inconvenient: it already manifests as duplicated dependencies and inconsistent installations across the community, exposing a gap that dependency management has yet to close.
In this study, we set out to study the intricacies of the dependencies contained in the agent skills (mixed skill-package-service dependencies) through introducing \emph{Agent Skill Supply Chains {\rm(}ASSCs{\rm)}}, expecting to help close this gap.
Borrowing the idea of \emph{Software Bill of Materials {\rm(}SBOMs{\rm)}}, we design SkillDepAnalyzer (\tool{}) to capture natural-language dependency evidence and model skills as dependency-bearing artifacts.
On SKILL-DEP benchmark, \tool{} recovers skill metadata and whole dependency graphs accurately and comprehensively, substantially outperforming an LLM-based baseline and package-centric SBOM tools. 
Applying \tool{} to over 1.43 million skills, we obtain the ASSCs and explore their structural diversity and security signals.
We find four structural patterns: skill metadata is activation-ready but governance-poor; dependency graphs span skill, package, and service dependencies with concentrated reuse; recursive skill reuse expands dependency graphs and creates hidden package dependency inventory; and skill dependency clusters are formed around related workflows.
We also find that inspecting a skill document alone is insufficient because security-relevant signals may hide in its dependencies. By analyzing ASSCs, we identify and report known malicious skills persisting in ASSCs to their developers.
Based on these findings, we recommend typed dependency manifests, first-class dependency-cluster management, risk-warning audit commands for skill infrastructure maintainers (e.g., developers of skill package managers and maintainers of skill registries), and lockfile-like records for skill developers.
\end{abstract}


\section{Introduction}

Agent skills encapsulate reusable operational knowledge that enables Large Language Model (LLM) agents to perform specialized tasks~\cite{anthropicSkillsDocs,openaiCodexSkills,openaiSkillsDocs}. A typical skill packages front matter, natural-language instructions, and code scripts, which together specify how such tasks are performed. 
To date, the number of publicly available skills has reached $1.43$ million, representing a ninefold increase in just three months~\cite{skillsmpMarketplace}. 
These skills span diverse domains, including networking, finance, data analysis, and software development. 
To avoid reinventing existing capabilities, developers increasingly compose new skills by reusing existing skills, software packages, and external services. Such widespread reuse accelerates skill development, but it also introduces layered \textbf{dependencies}, transforming skills from isolated files into dependency-bearing artifacts.

However, existing dependency-management mechanisms have not kept pace with this emerging reuse practice. Rather than being explicitly declared, skill dependencies are implicitly scattered across metadata, instructions, and scripts. Consequently, developers cannot reliably identify a skill's dependencies, determine their versions, or trace their provenance.

Recent community reports show that this opacity already causes practical dependency-management problems. 
A discussion in the Claude Code community reports that skills are distributed as copy-in snapshots, causing their upstream repositories, paths, and version information to be lost after submission~\cite{anthropicSkillsExternalRepo796}. As a result, provenance tracking and reliable updates become difficult. 
Another issue in the Claudekit community reports inconsistent handling of package dependencies required by skills, where different package managers are invoked under different circumstances, resulting in unpredictable dependency management behavior~\cite{claudekitCliIssue871}. 
Together, these cases highlight that systematic mechanisms for managing skill dependencies are still lacking. 

We therefore propose \emph{Agent Skill Supply Chains {\rm(}ASSCs{\rm)}} to model mixed skill-package-service dependency structures. 
An ASSC is a directed dependency graph whose nodes represent skills, software packages, and external services, and whose edges represent dependency relationships. 
ASSCs explicitly represent previously implicit skill dependencies, making it possible to trace dependency provenance, maintain reused components, and audit transitive risks.

To the best of our knowledge, no existing approach can construct ASSCs. 
The closest techniques are traditional \emph{Software Bill of Materials {\rm(}SBOMs{\rm)}}~\cite{spdx301,cyclonedxOverview}, which recover dependencies from software manifests. 
However, this assumption breaks down for agent skills. 
On the one hand, skill dependencies are often expressed in natural language rather than in machine-readable manifests. 
On the other hand, skills involve dependency types beyond the scope of traditional software package analyzers, including skill-to-skill reuse and dependencies on external services (e.g., Model Context Protocol (MCP) servers).


To systematically construct ASSCs, we develop an automated skill dependency analysis tool, named \tool{}. 
First, \tool{} recovers candidates for direct dependencies from front matter and skill bodies. 
Second, \tool{} assesses the confidence of each candidate. Low-confidence candidates are retained as annotations to preserve traceability, while high-confidence candidates are classified by a typed scanner into three dependency channels---skills, packages, and service. 
Third, SDA recursively resolves skill and package dependencies, incrementally recovering transitive dependencies to construct an ASSC. 
Finally, SDA serializes the recovered ASSCs into Skill Bill of Materials (SkillBOM) documents, a skill-oriented representation based on the SBOM Intermediate Representation (SBOM-IR)~\cite{jia2025sit}. 
Notably, SkillBOM is tailored to skills while remaining compatible with existing SBOM standards. This compatibility enables SBOM-based toolchains to be reused for analyzing ASSCs.

With \tool{}, we can systematically analyze skill dependencies at scale, enabling a large-scale study of ASSCs. 
First, we construct the SKILL-DEP benchmark to validate \tool{}, evaluating its accuracy and comprehensiveness in recovering skill metadata and dependencies. 
Second, we apply the validated tool to 1.43 million skills to characterize the structural patterns of ASSCs. 
Finally, we examine how security-relevant skills, packages, and services are exposed through ASSCs. 
Accordingly, we formulate the following three research questions.

\textbf{RQ1 (Analyzer Evaluation).} How accurately and comprehensively can \tool{} analyze skill metadata and dependencies?

\textbf{RQ2 (ASSC Characteristics).} What structural patterns characterize ASSCs?

\textbf{RQ3 (Security Propagation).} How are security-relevant skills, packages, and services exposed through ASSCs?

To answer RQ1, we construct a human-labeled SKILL-DEP benchmark to evaluate  
\tool{}. 
For accuracy, \tool{} achieves an overall F1 score of 0.95 on the single-layer benchmark, outperforming all baselines across all three dependency categories, and achieves perfect accuracy (1.00) on metadata fields. For comprehensiveness, it achieves an F1 score of 0.95 on the multi-layer benchmark across skill dependency graphs, confirming its ability of incremental BOM construction. 

To answer RQ2, we apply \tool{} to 1.43 million skills, revealing four structural characteristics of ASSCs that motivate governance recommendations. 
First, skill metadata infrastructure is \textbf{activation-ready but governance-poor}. Names and descriptions are widely available, but dependency, license, and version declarations remain sparse. 
Name collisions further complicate large-scale skill identity resolution, as 58.73\% of skills have non-unique names. 
Second, ASSCs \textbf{exhibit multi-channel dependencies with highly concentrated reuse}. 
Skill dependencies span three channels: skills (8.92\%), software packages (15.48\%), and external services (22.25\%). 
Reuse is highly concentrated around a small set of skills and packages, suggesting that ASSCs are built upon a narrow reusable core. 
Third, ASSCs exhibit \textbf{recursive dependency expansion and hidden package inventories}. 
Skill reuse substantially multiplies a skill's package dependencies, introducing many transitive packages that are not explicitly declared by the skill itself. 
In npm and PyPI, 71.87\% and 73.33\% of packages, respectively, are inherited through skill reuse, hidden from direct skill declarations. 
Fourth, ASSCs form \textbf{dependency clusters}. 
Overall, 30.41\% of dependency-bearing root skills belong to dependency clusters, which commonly connect skills implementing different stages of the same workflow. These clusters naturally emerge as high-level units for dependency management and governance. 

To answer RQ3, we analyze how security patterns propagate through ASSCs. 
We find that security-relevant signals across skills, packages,
and services propagate reside beyond direct dependencies. 
Specifically, 60–78\% of security-relevant skill dependencies, 98.01\% of \texttt{axios} package dependencies, and 93.10\% of potentially vulnerable MCP service dependencies are inherited exclusively through transitive dependencies, remaining invisible to reviewers who inspect only the root skill. 
Leveraging ASSCs, we identify security-relevant issues in the wild, including copies of the malicious \texttt{clawhub1} skill in the \texttt{Demerzels-lab/elsamultiskillagent} repository, and report them to the corresponding developers. 

These results motivate governance suggestions for two groups. 
For skill infrastructure maintainers (e.g., developers of skill package managers and maintainers of skill registries), we recommend adopting typed multi-channel dependency manifests that distinguish skill, package, and service dependencies while recording their source and version information. 
We also recommend first-class dependency-cluster management, allowing tightly coupled dependency clusters to be declared, resolved, and audited as high-level units. 
In addition, skill package managers can provide risk-warning audit commands to report vulnerable transitive dependencies. 
For skill developers, we recommend maintaining lockfile-like records that preserve pinned versions and source repositories whenever skills reuse dependencies across repositories.

In summary, this paper makes the following contributions:

\begin{itemize}[leftmargin=*]
    \item We define \emph{Agent Skill Supply Chains {\rm(}ASSCs{\rm)}} as mixed skill-package-service dependency graphs, and propose a method to retrieve and analyze ASSCs.
    \item We develop \tool{}, an automated tool that analyzes skill dependencies and emits SkillBOM, a skill-oriented BOM representation. We construct the SKILL-DEP benchmark, on which \tool{} outperforms all package-centric SBOM generators and an LLM-based baseline.
    \item We apply \tool{} to 1.43 million skills to characterize ASSC structures, revealing activation-ready but governance-poor metadata, multi-channel and concentrated dependency graphs, dependency expansion and hidden package inventory, and dependency clusters.
    \item We analyze security propagation in ASSCs and show that inspecting only the root skill misses security-relevant signals that appear in transitive dependencies, including known malicious skills uncovered through ASSC inspection.
\end{itemize}

The remainder of the paper is organized as follows. 
Section~\ref{sec:bg_and_related} reviews background on agent skills, and related work on software supply-chain transparency, skill reuse, and skill security.
Section~\ref{sec:sda} presents \tool{}, the automated analyzer for recovering skill metadata and dependencies.
Section~\ref{sec:rq1} evaluates the accuracy and comprehensiveness of \tool{}.
Section~\ref{sec:rq2} applies \tool{} to 1.43 million skills and characterizes ASSCs.
Section~\ref{sec:rq3} analyzes how security-relevant signals propagate through ASSCs.
Section~\ref{sec:discussion} discusses suggestions for ASSC governance and threats to validity.
Section~\ref{sec:conclusion} concludes this paper.

\section{Background and Related Work}
\label{sec:bg_and_related}

\subsection{Background} 

Agent skill is typically an artifact centered on a \texttt{SKILL.md} file, with optional scripts, references, or other bundled resources loaded as needed~\cite{anthropicSkillsDocs,openaiCodexSkills}. Prior work characterizes a skill document as a three-part artifact: YAML front matter, natural-language instructions, and code scripts~\cite{zhu2026skillclone}. These parts play different roles. Front matter usually describes skill metadata information such as name and description; the natural-language instructions explain when and how the agent should use the skill; code scripts can encode concrete operations such as package installation, API calls, or helper-script execution.

Unlike conventional software packages, skills do not provide a standardized dependency declaration mechanism. As a result, dependency evidence may be scattered across any of these three parts. Front matter may include dependency-like fields such as \texttt{dependencies} or \texttt{requires}; natural-language instructions may mention required packages, reused skills, or MCP servers; and code scripts may contain installation commands, imports, API clients, or configuration snippets.

Therefore, a skill cannot be analyzed only as a plain-text prompt or as a conventional package manifest. Dependency analysis must instead collect evidence across metadata, instructions, and code, while distinguishing whether each dependency points to a package, another skill, or an external service.

\subsection{Related Work}

\subsubsection{Software Supply Chain Transparency}

Software supply chain already has mature transparency formats and tools for conventional software artifacts. SBOM standards, such as SPDX~\cite{spdx301} and CycloneDX~\cite{cyclonedxOverview}, define relationships and metadata for software packages. Generators, such as Syft~\cite{syft}, Cdxgen~\cite{cdxgen}, ScanCode~\cite{scancode}, ORT~\cite{ort}, and Microsoft sbom-tool~\cite{microsoftSbomTool}, support automated supply chain visibility by extracting package dependencies from manifests, lockfiles, and build metadata. 
These SBOM standards and tools are useful for package-centered software supply chains, but they do not match skills. Existing SBOMs do not treat skills as independent components or represent skill-specific relations such as skill-to-skill reuse. Existing generators also assume package-oriented inputs, so they cannot recover these relations from \texttt{SKILL.md} metadata, natural-language instructions, and code. This gap motivates both our new skill dependency analyzer \tool{} and our new output representation SkillBOM.

\subsubsection{Skill Reuse}

Recent work frames agent skills as reusable artifacts rather than isolated prompts. OpenAI's skill documentation describes skills as reusable extension points for agent workflows~\cite{openaiSkillsDocs}. SkillClone treats agent skills as multi-channel artifacts and detects clone relationships across YAML metadata, natural-language instructions, and code scripts~\cite{zhu2026skillclone}. It shows that skill clones can propagate quality and security issues. These studies establish that skills are reused at scale and that reuse can carry hidden maintenance and security costs. Our work studies a different relation. Instead of detecting whether two skills are clones, we reconstruct dependency links among skills, packages, and services. This dependency graph allows us to analyze ASSC structure and security-relevant patterns.

\subsubsection{Skill Security}

Skill-specific security work studies risks inside skills. Registry-level empirical work analyzes third-party skill registries and reports confirmed malicious skills with behaviors such as credential theft, remote-code execution, and adversarial instructions~\cite{liu2026maliciousskills}. Marketplace-scale studies report vulnerability patterns including prompt injection, data exfiltration, privilege escalation, and supply-chain risks~\cite{liu2026agentskillswild}. Benchmark work further measures whether agents follow prompt injections placed inside skill files~\cite{schmotz2026skillinject}. Red-teaming work shows that non-malicious skills can still be exploited through adversarial prompting~\cite{duan2026skillattack}. 
Our work focuses on a different perspective. We analyze how security-relevant skills, packages, and services become reachable through ASSCs.

\section{SkillDepAnalyzer}
\label{sec:sda}

We introduce the automated SkillDepAnalyzer tool (SDA) that we develop in this section. SDA is the infrastructure of this study.

\subsection{Overview}

\begin{figure*}[t]
\centering
\includegraphics[width=\textwidth]{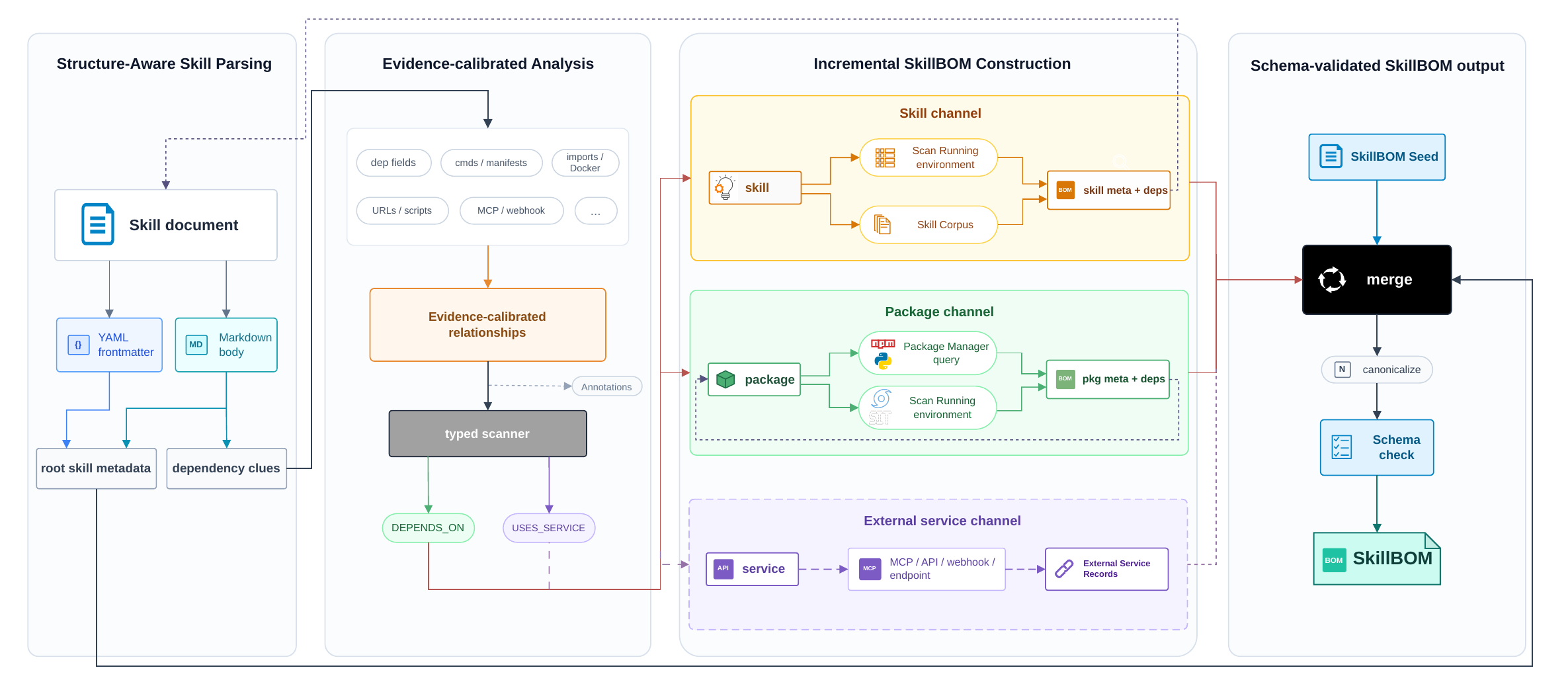}
\Description{Pipeline diagram showing skill document input, YAML front matter and skill body parsing, typed evidence scanning, evidence-calibrated relationships, package, skill, and external service construction channels, merge, canonicalization, schema check, and final SkillBOM output.}
\vspace{-1.5em}
\caption{\tool{} generation workflow.}
\vspace{-1em}
\label{fig:pipeline}
\end{figure*}

\tool{} converts a skill document into a structured SkillBOM through the pipeline shown in Figure~\ref{fig:pipeline}. 
First, given a skill document, \tool{} extracts root-skill metadata and dependency clues from the skill content. 
Second, \tool{} assesses the confidence of each candidate and classifies confirmed dependencies into three channels: skills, packages, and external services.
Third, \tool{} recursively resolves referenced skills and packages and records external services to construct an ASSC.
Finally, the recovered ASSC is serialized into a structured SkillBOM.

\subsection{Structure-Aware Skill Parsing}

\tool{} first separates each skill document into front matter and the skill body. Front matter is usually written in YAML, so \tool{} analyzes it structurally and recovers root-skill information such as \texttt{name}, \texttt{version}, and \texttt{description}. These fields are treated as the highest-priority evidence for the root skill identity. If front matter is absent or does not provide enough metadata, \tool{} falls back to the skill body. The analyzer uses metadata keyword matching and dependency pattern matching to complement root-skill metadata and collect dependency clues from dependency fields, commands, or natural-language mentions.

\subsection{Evidence-Calibrated Dependency Analysis}

All collected clues form an initial pool of candidate dependency evidence. Since the same artifact may appear in different contexts with different semantics (e.g., a package may be required by an installation command or merely shown in an example), \tool{} evaluates each candidate together with its surrounding context and supporting evidence. Confirmed candidates are then classified into three dependency channels—packages, skills, and external services—while ambiguous candidates are retained as annotations.

Package dependencies are analyzed from package-specific evidence patterns, including package managers, installation commands, manifest entries, and Docker snippets. Strong evidence, such as installation commands and manifest entries, directly establishes package dependency edges. For weaker package mentions, such as package names extracted from textual descriptions rather than explicit installation instructions, \tool{} queries the corresponding package registry and creates a dependency edge only when an exact package match is found and the surrounding context supports a dependency interpretation. Otherwise, the candidate is retained as an annotation so that examples and troubleshooting text do not enter the dependency graph.

Skill dependencies require more than name matching because skills are frequently cloned or renamed. \tool{} therefore resolves candidate skills using both names and identity information, including repositories and paths. Matches confirmed by repository or path evidence are recorded as skill dependencies, whereas candidates identified only by skill names, without a corresponding match in the running environment or skill dataset, are retained as annotations.

External services follow a similar calibration path. A skill may assume an MCP server, API endpoint, or webhook. \tool{} extracts service cues from natural-language instructions or code scripts and records observable service-use evidence as service records.

Overall, this stage produces the root skill metadata and its first-layer evidence-calibrated dependency, covering direct package dependencies, direct skill dependencies, service records, and annotations. Only high-confidence edges are treated as true dependencies for downstream analyses, while lower-confidence matches are retained as annotations.

\subsection{Incremental BOM Construction}

This stage takes the root skill metadata and first-layer dependency from the previous stage as input. \tool{} first constructs a seed SkillBOM for the root skill. It then expands the seed through two dependency channels: the package channel, which enriches package metadata and package-level transitive dependencies, and the skill channel, which imports matched dependency skills and their generated SkillBOMs. Because service records mainly describe external services, \tool{} does not expand them for transitive dependencies. Transitive facts about packages and skills then enter the final result through recursively resolved package and skill dependency records.

The package channel expands direct package dependencies into package-level dependency graphs. \tool{} prioritizes evidence from the local running environment when available. \tool{} scans the environment to recover installed package metadata and transitive dependencies for each package, then uses these records to construct the package dependency graph. If no running environment is available, \tool{} falls back to registry-level resolution. When the package registry or package manager can be inferred, \tool{} queries the corresponding package manager or registry to recover package metadata and dependencies. Through local running environments and package registries, \tool{} queries and verifies package information, then constructs a comprehensive package dependency graph.

The skill channel starts from the direct skill dependencies. For each dependency skill, \tool{} first searches the local running environment and then the skill dataset. Candidate skills are matched using multiple identity fields, including skill name, repository, path, owner, and repository stars, with earlier fields assigned higher priority. Thus, matches confirmed by skill name, repository, and path are considered more reliable than those supported only by weaker evidence, such as repository stars. If no exact skill-name match is found, \tool{} further considers prefix and suffix name matches and ranks the resulting candidates using the same ranking strategy. The highest-ranked candidate is selected as the resolved dependency if its score exceeds the confidence threshold; otherwise, the candidate is retained as an annotation.

After selecting a matched skill, \tool{} analyzes that skill document with the same pipeline to get its metadata and skill-package-service graphs and imports its generated SkillBOM into the current root result. This import is recursive and cycle-aware. Because of skill dependency clusters, \tool{} keeps a visited-skill list: once a skill is added to the graph, later visits to the same skill stop expansion. The merge step canonicalizes duplicates and preserves relationship evidence.

Overall, this stage expands a root-local SkillBOM seed into a comprehensive dependency graph while preserving how each imported record entered the result.

\subsection{Structured Schema-Validated SkillBOM Output}

In this stage, \tool{} converts the dependency graph from the previous stage into a structured schema-validated SkillBOM that represents all the skills, packages, and external services in one machine-readable object. Before emission, \tool{} canonicalizes duplicate components, preserves relationship evidence and annotations, and validates the result against the SkillBOM schema to ensure a unified representation.

SkillBOM is a skill-oriented representation built on SBOM IR~\cite{jia2025sit}. It treats a skill as a first-class component, assigns UUID-based skill identifiers, and records skill-specific component types. Its relationship model also captures skill-oriented dependencies, such as external-service-use relationships like \texttt{USES\_SERVICE}. Because SkillBOM is built on SBOM-IR, it remains compatible with international SBOM standards such as SPDX~\cite{spdx301} and CycloneDX~\cite{cyclonedxOverview}. SkillBOM can therefore import and export SBOM documents and reuse existing SBOM-based toolchains for ASSC analysis.

\section{RQ1: Analyzer Evaluation}
\label{sec:rq1}

RQ1 examines whether \tool{} can serve as a trustworthy measurement instrument for ASSC analysis. We evaluate \tool{} using the SKILL-DEP benchmark, which contains a single-layer benchmark for metadata and direct-dependency recovery and a multi-layer benchmark for whole skill dependency-graph construction. The single-layer benchmark measures \tool{}'s accuracy against metadata and direct dependencies, while the multi-layer benchmark measures its comprehensiveness in constructing complete skill dependency graphs.

\subsection{SKILL-DEP Benchmark Construction}

\textbf{Single-layer benchmark.}
The single-layer benchmark is built from a stratified sample of 500 skill documents. Each sampled unit is one root skill document, and each unit is annotated with two gold targets: the root-skill metadata and the set of package, skill, and service-use dependencies expressed in that document. We use open annotation throughout benchmark construction. Because dependency evidence is sparse and unevenly distributed, we stratify the sample by dependency-evidence type to ensure coverage of package dependency evidence, skill dependency evidence, and service-use evidence.

We calibrate the annotation guideline on a 50-skill pilot. Two software-engineering researchers with skill-usage experience independently extract dependency records, compare disagreements, refine the definitions, and freeze the guideline before full annotation. The final guideline defines dependency inclusion criteria, the boundaries among package, skill, and service dependencies, target canonicalization rules, and recorded relationship types. Each record contains the source skill, target kind, canonicalized target name, relationship type, and evidence location; skill-level metadata such as name, repository, path, and license were also recorded when available.

The same two researchers then independently annotate the 500-skill benchmark under the frozen guideline. We align extracted records by source skill, target kind, canonicalized target name, and relationship type. Before adjudication, the two annotators extract 1,583 and 1,590 dependency records, with 1,581 matched records. The remaining 11 disputed records are resolved through consensus discussion: 5 are retained as valid dependencies, and 6 are rejected as non-dependencies or out-of-scope cases, such as example-only mentions. The matched records plus the retained disputed records form the frozen adjudicated reference set.

\textbf{Multi-layer benchmark.}
To validate the whole dependency graph construction, we sample 100 skills and apply the same frozen guideline and annotation procedure. These skills focus on skill graph construction: it contains 80 graphs of depth 3, 18 of depth 4, and 2 of depth 5. We use it to test whether \tool{}'s incremental construction extracts multi-layer skill dependencies consistently.

\begin{table}[tbp]
\centering
\caption{Data characteristics of the SKILL-DEP benchmark.}
\label{tab:benchmark-characteristics}
\begin{tabular*}{\columnwidth}{@{\extracolsep{\fill}}llcc}
\toprule
Benchmark part & Benchmark characteristic & Count & Proportion \\
\midrule
\multirow{5}{*}{Single-layer}
& Root skill documents & 500 & - \\
& Package dependencies & 436 & 27.5\% \\
& Skill dependencies & 708 & 44.6\% \\
& Service-use dependencies & 442 & 27.9\% \\
\midrule
\multirow{4}{*}{Multi-layer}
& Root skill documents & 100 & - \\
& Depth-3 skill graphs & 80 & 80.0\% \\
& Depth-4 skill graphs & 18 & 18.0\% \\
& Depth-5 skill graphs & 2 & 2.0\% \\
\bottomrule
\end{tabular*}
\end{table}

Table~\ref{tab:benchmark-characteristics} reports the combined benchmark characteristics after adjudication. For dependency scoring in both single-layer and multi-layer benchmarks, each dependency row is keyed by source skill, target kind, canonicalized target name, and relationship type. Thus, package, skill, and service targets are evaluated under a unified dependency-extraction target while preserving their type distinction. Root-level metadata is scored separately at the sampled-skill level by comparing predicted fields, such as the skill name, source repository, and declared license.

\subsection{Experimental Setup}

\textbf{Baselines.}
We compare \tool{} with two kinds of baselines. 
The first is package-centric SBOM generator baselines. We select widely used open-source tools from GitHub repositories tagged with the topic \texttt{sbom}. We include tools with at least 2,000 stars and explicit SBOM generation capability: Syft~\cite{syft}, ScanCode~\cite{scancode}, ORT~\cite{ort}, and Microsoft sbom-tool~\cite{microsoftSbomTool}. In addition, we include Cdxgen~\cite{cdxgen}, the most-starred SBOM generator maintained by the CycloneDX community, to cover a generator developed within a major SBOM standards community.

The second baseline is an LLM-based extractor, which tests whether semantic extraction alone can recover dependency records and metadata. Considering the practical balance between model capability and evaluation cost for running the full benchmark, we use DeepSeek-v4-pro with a fixed prompt that instructs the model to act as a software-engineering researcher with skill-usage experience and return only valid JSON. The prompt includes instructions on what dependencies to extract and how to format the output. We use temperature 0 to minimize sampling variance. 
LLM-based baseline complements package-centric SBOM generators, which cannot extract skill and service-use dependencies, and provides a direct comparison point for assessing whether \tool{}'s typed analysis improves over an LLM-based semantic extractor.

\textbf{Metrics.}
For dependency extraction in both single-layer and multi-layer benchmarks, we report precision, recall, and F1 score. For metadata extraction, we report field-level accuracy.

\subsection{Evaluation Results}

\subsubsection{Accuracy}

\begin{table*}[tbp]
\centering
\caption{Evaluation of Dependency Extraction on the Single-layer SKILL-DEP Benchmark.}
\label{tab:rq1-dependency-accuracy}
\begin{tabular*}{\textwidth}{@{\extracolsep{\fill}}ccccccccccccc}
\toprule
& \multicolumn{3}{c}{Overall} & \multicolumn{3}{c}{Package} & \multicolumn{3}{c}{Skill} & \multicolumn{3}{c}{Service} \\
\cmidrule(lr){2-4}\cmidrule(lr){5-7}\cmidrule(lr){8-10}\cmidrule(lr){11-13}
Method & P & R & F1 & P & R & F1 & P & R & F1 & P & R & F1 \\
\midrule
Cdxgen & 0.33 & 0.14 & 0.19 & 0.33 & 0.50 & 0.40 & - & - & - & - & - & - \\
Syft & 0.38 & 0.14 & 0.20 & 0.38 & 0.50 & 0.43 & - & - & - & - & - & - \\
ScanCode & 0.39 & 0.18 & 0.25 & 0.39 & 0.66 & 0.49 & - & - & - & - & - & - \\
ORT & 0.52 & 0.11 & 0.18 & 0.52 & 0.39 & 0.45 & - & - & - & - & - & - \\
Microsoft sbom-tool & 0.60 & 0.15 & 0.24 & 0.60 & 0.56 & 0.58 & - & - & - & - & - & - \\
DeepSeek & 0.48 & 0.55 & 0.52 & 0.43 & 0.66 & 0.52 & 0.56 & 0.50 & 0.53 & 0.45 & 0.54 & 0.49 \\
\tool{} & 0.92 & 0.98 & 0.95 & 0.89 & 0.97 & 0.93 & 0.95 & 0.98 & 0.96 & 0.91 & 0.97 & 0.94 \\
\bottomrule
\end{tabular*}
\end{table*}

\textbf{Dependency extraction.} 
Table~\ref{tab:rq1-dependency-accuracy} shows that \tool{} achieves the highest 0.95 overall F1, substantially exceeding every package-centric SBOM generator and DeepSeek, even though DeepSeek has semantic extraction capability. 

For package dependencies alone, \tool{} reaches 0.93 F1, outperforming the best package-centric SBOM generator. This shows that skill dependency extraction is not equivalent to conventional SBOM generation over source code. Skills mix YAML metadata, natural-language instructions, and code scripts, so dependency evidence is more scattered and less standardized. 
\tool{} performs better because it is customized for this mixed-document structure. Typed evidence scanners capture channel-specific patterns, while contextual calibration filters examples, templates, and identifier-like mentions without dependency intent. By contrast, DeepSeek mainly suffers from weak boundary control: it over-predicts semantically related but non-required items and misses implicit skill dependencies expressed through paths or workflow conventions.

The per-category results show where this advantage comes from and where errors remain. Package dependencies are hardest with 0.93 F1 because examples and command templates can resemble real package requirements, leading to false positives. Skill dependencies perform best with 0.96 F1 because many edges contain explicit names or source cues. Service-use dependencies sit between them with 0.94 F1, as MCP/API/webhook cues are visible but less standardized than skill-file references.

\begin{table}[tbp]
\centering
\caption{Evaluation of Metadata Extraction on the Single-layer SKILL-DEP Benchmark.}
\label{tab:rq1-metadata-accuracy}
\begin{tabular*}{0.8\columnwidth}{@{\extracolsep{\fill}}ccccc}
\toprule
Method & Name & Repo & Path & License \\
\midrule
LLM-based & 1.00 & 0.99 & 0.99 & 1.00 \\
\tool{} & 1.00 & 1.00 & 1.00 & 1.00 \\
\bottomrule
\end{tabular*}
\end{table}

\textbf{Metadata extraction.}
As shown in Table~\ref{tab:rq1-metadata-accuracy}, both systems recover metadata well, but \tool{} is stronger on the repository and path fields that matter most for traceability. This further demonstrates that \tool{}'s skill-oriented design is important for accurate metadata extraction.

\subsubsection{Comprehensiveness}
On the multi-layer benchmark, \tool{} achieves precision 0.98, recall 0.93, and F1 0.95. Since no baseline supports multi-layer dependency extraction, we evaluate \tool{} against the human-annotated benchmark directly. The result shows that \tool{}'s incremental BOM construction maintains high accuracy across multi-layer skill dependency graphs, demonstrating its comprehensiveness in constructing whole skill dependency graphs.

\begin{rqanswerbox}
\textbf{Answering RQ1:} \tool{} analyzes skill metadata and dependencies accurately and comprehensively. It achieves 0.95 overall dependency F1 on the single-layer benchmark, outperforming all baselines across the three dependency categories, and 1.00 accuracy on all metadata fields. On the multi-layer benchmark, it achieves 0.95 F1, confirming its ability to construct whole dependency graphs.
\end{rqanswerbox}

\section{RQ2: ASSC Characteristics}
\label{sec:rq2}

In this section, we construct a large-scale skill dataset from the SkillsMP registry and apply \tool{} to build ASSCs, which we analyze from two perspectives: metadata infrastructure and dependency graphs.

\subsection{Skill Dataset}

We build the skill dataset from the SkillsMP registry~\cite{skillsmpMarketplace}. SkillsMP indexes all public skill files from GitHub and organizes them by keywords, creators, and source repositories, making it a suitable entry point for large-scale skill analysis. On June 6, 2026, SkillsMP listed 1,640,440 skills. After capturing a snapshot of the registry that day, we retrieved each skill's GitHub URL and successfully downloaded 1,434,046 GitHub-backed records, covering about 87.4\% of the listed skills. The remaining records could not be obtained because their GitHub links were inaccessible, their repositories had become private, or their source content was otherwise unavailable at crawl time. The analyzed snapshot contains repository/path metadata and SHA-256 content hashes. We retain these identities rather than deduplicate the 0.52\% hash-identical records, since identical skill files may appear under different names or contexts that affect downstream dependency resolution.

\subsection{Metadata Infrastructure}
\label{sec:rq2-metadata}

Skills carry enough front matter to support agent activation but far too little to support supply-chain governance. Across 1,434,046 skills, front matter is present in 99.55\% of skills; \texttt{name} appears in 99.49\% and \texttt{description} in 99.52\%. In contrast, \texttt{license} appears in only 11.25\% of skills and \texttt{version} in only 20.12\%. Fields that could carry dependency-like declarations (e.g., \texttt{dependencies}, \texttt{requires}) appear in only 1.40\% of skills, while over 30\% of skills actually carry package, skill, or service-use dependencies. 
The contrast is structural rather than incidental. The skill format is designed for agent scheduling: name and description tell an agent when to invoke a skill. Fields that tell a human reviewer or an automated scanner \emph{what the skill depends on} are rarely standardized. 

Worse still, even the metadata fields that do exist are fragile as identifiers. Effective names, including those derived from front matter \texttt{name} or the parent directory slug, collide across repositories: 58.73\% of skills share their effective name with at least one other record. Among these colliding names, 88.76\% span multiple repositories and 99.31\% correspond to different content hashes. This is not a benign scenario where identical skills appear across multiple locations, but a harder one involving different skills sharing identical or similar names. Descriptions are similarly unreliable: 25.88\% of skills share a normalized description with at least one other record, and 11.57\% have a front matter \texttt{name} that differs from their parent path slug.

Without well-managed and standardized metadata declarations, dependency resolution, provenance tracking, and risk notification all become rather tricky. The community has already begun discussing this kind of problem~\cite{opencodesdlcwizardIssue26}. Current proposals remain limited to non-deterministic agent-based inference at skill-use time to fill in likely metadata. This problem has become a measurable pattern of the metadata infrastructure that any skill governance system must address.

\subsection{Dependency Graph}
\label{sec:rq2-graph}

\textbf{Multi-channel dependencies.} Skill dependencies distinguish three dependency channels: skill dependencies, package dependencies, and service-use dependencies.

\begin{table}[t]
\centering
\caption{Dependency Channels in Skills.}
\label{tab:dependency-dimensions}
\begin{tabular}{ccc}
\toprule
Channels & Roots & Proportion \\
\midrule
Direct skill dependency & 127,891 & 8.92\% \\
Direct package dependency & 221,925 & 15.48\% \\
Direct service-use dependency & 319,013 & 22.25\% \\
Any of the three & 524,802 & 36.60\% \\
All three & 11,041 & 0.77\% \\
\bottomrule
\end{tabular}
\end{table}

Table~\ref{tab:dependency-dimensions} reveals that the three dependency channels capture largely distinct populations---only 0.77\% of skills carry evidence in all three, and the combined surface of 36.60\% is nearly four times the skill-only channel alone. Notably, skill dependencies---the channel that has drawn the most attention from recent studies~\cite{bhardwaj2026formal,liu2026agentskillswild,zhu2026skillclone}---account for the smallest fraction of dependency-bearing skills (8.92\%), while package and service-use dependencies together cover much more ground. This gap suggests that the current research focus on skill reuse underestimates the broader supply-chain surface that skills actually participate in.

\textbf{Concentration.} The dependency graph concentrates around a small set of heavily reused targets. Under the highest confidence repo/path-context identity filter, skill-to-skill normalized Gini (a size-normalized measure of dependency concentration, ranging from 0 to 1)~\cite{decan2019empirical} reaches 0.925 and the top-20 skill targets concentrate 14.12\% of all skill edges. Package targets are similarly concentrated: normalized Gini reaches 0.944, with the top-20 packages accounting for roughly 8.8\% of all package edges. For reference, both values exceed npm's normalized dependency Gini of 0.87~\cite{decan2019empirical}---a registry already known for its heavily concentrated dependency distribution. That ASSCs already concentrate around a small set of targets despite their recent emergence suggests that skill reuse is converging rapidly around a few widely adopted hubs, rather than spreading evenly across the supply chain.

\begin{table}[t]
\centering
\caption{Total and package amplification in ASSC.}
\label{tab:amplification}
\begin{tabular}{ccccc}
\toprule
Amplification & p50 & p90 & p99 & Max \\
\midrule
Total & 0.5 & 23.0 & 130.5 & 979.0 \\
Skill & 0.0 & 4.1 & 34.0 & 347.0 \\
Package & 0.0 & 47.0 & 350.0 & 1754.0 \\
Service & 0.0 & 5.0 & 63.5 & 194.0 \\
\bottomrule
\end{tabular}
\end{table}

\textbf{Dependency expansion and hidden package inventory.}
We use the \emph{dependency amplification} factor~\cite{arafat2025deep}, the ratio of transitive to direct dependencies, to measure how much a root's dependencies expand after recursive skill imports. Table~\ref{tab:amplification} shows a highly skewed pattern. The median total amplification is only 0.5, but the tail is large: p99 reaches 130.5$\times$ overall, with maxima of 347$\times$ for skills, 1,754$\times$ for packages, and 194$\times$ for services.

A top case explains the mechanism. \texttt{windows-95-web-designer} declares no package dependency and only three skill dependencies, yet its dependency graph imports 1,754 packages and 1,938 total components---645$\times$ amplification. Manual inspection of this and other high-amplification roots shows that such roots often act as workflow orchestrators: they name a few entry-point skills to execute concrete functions, while those skills recursively import their own dependencies.

This expansion creates a measurable governance consequence: hidden package inventory. Among dependency-bearing skills, 22.42\% gain packages only through reused skills, making those packages invisible at the root layer. A concrete example is \texttt{npm/rimraf}: 1,495 roots declare it directly, but 5,160 additional roots inherit it through skill reuse, meaning that for every skill that explicitly depends on \texttt{rimraf}, more than four others carry it without naming it.

This hidden package inventory couples ASSCs to software supply chains. Among npm package exposures, 71.87\% are inherited through skill reuse rather than directly declared; for PyPI, the share reaches 73.33\%. Thus, the ASSC does not form an isolated supply-chain island. It sits on top of npm, PyPI, and other supply chains, with skill reuse acting as the bridge that carries package inventory across skill boundaries.

\textbf{Dependency clusters.} The amplification extremes described above are not isolated abnormalities. They point to a structural phenomenon in ASSC: the \textbf{skill dependency cluster}. Skills often reference one another bidirectionally, forming mutual-dependency clusters that are conceptually closer to npm's peer dependencies, where packages declare compatibility with one another. The data bear this out. The skill dependency graph is not a tree: 30.41\% of root skills with dependencies contain at least one skill in a cycle, and 30.03\% have convergent downstream nodes. Among these, the most distinctive structure is a dependency network containing at least one strongly connected component whose members are drawn into a single mutually-referencing cluster. When any skill in such a cluster is installed, its dependencies pull in the others, making the entire cluster a high-level bundled unit.

A concrete illustration comes from \texttt{microsoft-defender-endpoint}, a Defender Advanced Hunting skill from \texttt{OpenTideHQ/AgentTide}. The root skill delegates generic detection-rule design to \texttt{detection-engineering}, which in turn is mutually linked with \texttt{microsoft-sentinel} and \texttt{threat-hunting}. These dependency skills perform specific functions and return the corresponding results, such as platform-specific detection guidance and threat-hunting practice. The result is a small dependency cluster whose members must be understood as a coupled workflow unit. 
The same pattern appears at larger scale in the \texttt{bitcoin-psbt} suite from \texttt{claude-dev-suite/claude-dev-suite}, which organizes 140 mutually referencing skills around Bitcoin custody knowledge, from timelocks and signatures to Taproot scripts, wallet descriptors, and vault protection.

When skills form dependency clusters like these, traditional governance tasks break down. Reviewing or auditing a single skill in isolation is not enough, because its dependencies and operational surface are extended by the cluster, not by its own \texttt{SKILL.md}. Patching or removing one member of the cluster may require coordinated changes across all members, and a vulnerability or risky instruction in any single skill propagates to every skill that depends on any entry point of the cluster.

\begin{rqanswerbox}
\textbf{Answering RQ2:}

\textbf{(1) Activation-ready but governance-poor metadata:} Front matter is present in 99.55\% of skills, but dependency-like fields appear in only 1.40\%. Meanwhile, 58.73\% of skill names collide, indicating a fragile metadata infrastructure.

\textbf{(2) Multi-channel and concentrated dependency graphs:} ASSCs span skill, package, and service dependencies that expose different parts of ASSCs, while reuse concentrates around a small set of skills and packages with normalized Gini of 0.925 and 0.944.

\textbf{(3) Dependency expansion and hidden package inventory:} Recursive skill reuse expands dependency graphs: p99 package amplification reaches 350$\times$, 22.42\% of dependency-bearing skills gain packages only through reused skills, and inherited package shares reach 71.87\% for npm and 73.33\% for PyPI. ASSC governance must inspect graph-level package exposure.

\textbf{(4) Dependency clusters:} The skill dependency graph contains clusters: 30.41\% of dependency-bearing roots include a cycle. These structures often connect skills that describe different parts of the same workflow, making the dependency cluster a high-level governance unit.

\end{rqanswerbox}

\section{RQ3: Security Propagation}
\label{sec:rq3}

In this section, we investigate how security-relevant signals from skills, packages, and services propagate through ASSC dependency graph. To do so, we collect security-relevant signals through two complementary approaches.

\begin{itemize}[leftmargin=*]
    \item \textbf{Indicators from public threat reports}

    Skills: malicious skill labels reported by Koi~\cite{koiClawHavoc}, Snyk~\cite{snykToxicSkills,clawdhubCampaign}, and Antiy~\cite{antiyClawHavoc}; 
    Packages: packages from documented supply-chain attacks, including Cline~\cite{clineCompromise}, Nx~\cite{nxS1ngularity} and axios~\cite{axiosCompromise};
    Services: MCP services with published CVEs or malicious MCP services, including mcp-remote~\cite{mcpRemoteCVE}, postmark-mcp~\cite{postmarkMcpMalicious}, MCP Inspector~\cite{inspectorCVE}, and figma-developer-mcp~\cite{figmaMcpCVE}.

    \item \textbf{Regex-based security patterns}
    
    Skills: six families of skill security-relevant signals concluded from academic papers and public reports~\cite{liu2026maliciousskills,skillThreatModeling}: remote payload execution, dangerous code patterns, prompt injection, credential exfiltration, persistence and backdoor commands, and secret exposure.
    Services: five families of service security-relevant signals concluded from official documentation and public reports~\cite{mcpSecurityBestPractices,owaspMcpToolPoisoning}: code repository and issue tracker authority, local resource and browser authority, collaboration and messaging authority, data backend authority, and mechanism-only MCP attack patterns.
\end{itemize}

Together, the two approaches identify potential security-relevant indicators. We scan the skill dataset and all generated SkillBOMs to observe how these indicators spread through the dependency graph. We then manually confirm cases such as the persisting \texttt{clawhub1} malicious skill and unpinned \texttt{axios} package.

\subsection{Skill Security Signals}
\label{sec:rq3-skill}

\begin{table}[t]
\centering
\caption{Propagation of skill security signals.}
\label{tab:rq3-skill}
\begin{tabular}{>{\raggedright\arraybackslash}p{0.23\columnwidth}ccc}
\toprule
Signal family & Root skills reached & Dependency-only & Proportion \\
\midrule
Publicly reported malicious skill labels & 194 & 26 & 13.40\% \\
Remote payload execution & 6,097 & 3,888 & 63.77\% \\
Dangerous code patterns & 3,342 & 2,621 & 78.43\% \\
Prompt injection & 1,413 & 913 & 64.61\% \\
Credential exfiltration & 1,189 & 812 & 68.29\% \\
Persistence / backdoor commands & 454 & 273 & 60.13\% \\
Secret exposure & 577 & 377 & 65.34\% \\
\bottomrule
\end{tabular}
\end{table}

Table~\ref{tab:rq3-skill} reports the propagation of skill-level security signals through the skill dependency graph. 
About 13.40\% of skills inherit publicly reported malicious signals purely through transitive dependencies. 
Notably, manual review identifies multiple cases where known malicious skills remain reachable through dependencies. For example, we find copies of \texttt{clawhub1} and \texttt{clawbhub}, which are originally reported as malicious under \texttt{zaycv}, in \texttt{Demerzels-lab/elsamultiskillagent}. Any skill in that repository that depends on the same collection can therefore transitively depend on these malicious skills. We report these issues to the corresponding developers.
For regex-based security families, 60--78\% of affected roots carry security-relevant patterns only through transitive dependencies. For example, \texttt{github-repo-management} by \texttt{jk-kim0/skills-jk} triggers remote-execution and shell-pipe patterns, and 44 of the 47 roots that reach it inherit the match purely through dependency. This shows why inspecting only the root skill can miss security-relevant patterns introduced by its dependencies.

\subsection{Package Security Signals}
\label{sec:rq3-package}

\begin{table}[t]
\centering
\caption{Propagation of package security signals.}
\label{tab:rq3-package}
\scriptsize
\begin{tabular}{cccc}
\toprule
Seed package & Root skills reached & Dependency-only & Proportion \\
\midrule
Axios & 3,413 & 3,345 & 98.01\% \\
Nx & 81 & 53 & 65.43\% \\
Clinejection payload & 38 & 17 & 44.74\% \\
Clinejection entry & 2 & 0 & 0.00\% \\
\bottomrule
\end{tabular}
\end{table}

Table~\ref{tab:rq3-package} reports the propagation of package-level security signals.
Package-level exposure is overwhelmingly driven by transitive dependencies: 98.01\% of roots carrying \texttt{axios} and 65.43\% carrying \texttt{nx} inherit them through their dependency graph. Manual inspection shows that the observed issue arises from install instructions without version pinning, causing whatever package version is locally served to be installed. Thus, root skills may remain exposed to packages with potential security issues because their inherited installation instructions are version-unstable.

\subsection{Service Security Signals}
\label{sec:rq3-service}

\begin{table}[t]
\centering
\caption{Propagation of service security signals.}
\label{tab:rq3-service}
\scriptsize
\begin{tabular}{>{\raggedright\arraybackslash}p{0.28\columnwidth}ccc}
\toprule
Service signal family & Root skills reached & Dependency-only & Proportion \\
\midrule
Publicly reported vulnerable MCP service hits & 29 & 27 & 93.10\% \\
Code repository / issue tracker authority & 3,566 & 2,600 & 72.91\% \\
Local resource / browser authority & 882 & 648 & 73.47\% \\
Collaboration / messaging authority & 303 & 161 & 53.14\% \\
Data backend authority & 218 & 157 & 72.02\% \\
Mechanism-only MCP attack patterns & 568 & 378 & 66.55\% \\
\bottomrule
\end{tabular}
\end{table}

Table~\ref{tab:rq3-service} reports the propagation of service-level security signals.
Reported vulnerable MCP services reach 29 roots, of which 27 (93.10\%) inherit them transitively. For example, \texttt{postmark-mcp}, a malicious MCP server documented by Snyk~\cite{postmarkMcpMalicious}, is still referenced by \texttt{aegis-research-lab}. Downstream skills in the same repository that depend on \texttt{aegis-research-lab} therefore inherit this exposure without ever naming the malicious service. Across broader service-authority signals, 53--73\% of affected roots are dependency-only. This means a skill that does not interact with a database or a browser can be exposed to those authority surfaces if any skill in its graph does. 
For instance, \texttt{activator-authoring-cli} inherits GitHub repository authority through a two-layer dependency path via \texttt{check-updates}, without ever naming the GitHub MCP server itself.

Taken together, the three channels tell the same story: vulnerable artifacts can enter ASSCs through any channel, persist long after discovery, and spread through the dependency graph to root skills that may be unaware that they authorize or use them.

\begin{rqanswerbox}
\textbf{Answering RQ3:}
Security-relevant signals enter ASSCs through skills, packages, and services. Malicious skills or services can persist through cloning, while unpinned install commands keep vulnerable packages reachable. The dependency graph amplifies this issue: dependency-only exposure reaches up to 98.01\%, so root-level inspection often misses exposures visible only in the full graph.
\end{rqanswerbox}

\section{Discussion}
\label{sec:discussion}

\subsection{Implications and Suggestions}

Based on our analysis of the structural characteristics of ASSCs and the patterns of security propagation, we outline suggestions and potential solutions for improving ASSC management. These suggestions are organized for two target groups: skill infrastructure maintainers (e.g., developers of skill package managers and maintainers of skill registries) and skill developers. 

\textbf{For skill infrastructure maintainers.} Early infrastructure for managing skills already exists. Industrial tools such as Vercel Labs' Skills CLI and the skills.sh index support skill management operations such as installation, update, and removal through \texttt{npx skills}~\cite{vercelLabsSkills,skillsSh}; research tools such as Skilldex propose conformance scoring, and registry APIs for skills~\cite{saha2026skilldex}. However, existing skill package managers still lack systematic mechanisms for dependency management and for warning users about vulnerable dependencies. We therefore recommend three improvements.

First, registries may require a typed dependency manifest. Instead of treating all references as generic related skills, the manifest can distinguish \texttt{skillDependencies}, \texttt{packageDependencies}, and \texttt{serviceDependencies}, and record metadata, such as source and version, for each dependency.

Second, skill package managers may introduce the concept of \emph{skill dependency clusters} into dependency management. The npm's experience with peer dependencies illustrates the importance of treating cluster-like relations as first-class dependency-management objects to avoid the version-governance failures that npm 4--6 experienced under weak peer-dependency enforcement~\cite{installPeerDependencies,npmV7Changes}. This urgency calls for governance at both the declaration and installation levels. At the declaration level, dependency manifests can include a \texttt{skillDependencyCluster} field to group related skills in the same cluster as a shared dependency unit, including optional dependencies for specific workflows. At the installation level, skill package managers can follow npm v7's practice of automatic resolution with conflict blocking~\cite{npmV7Changes}. When a consumer installs a skill, the manager should verify cluster compatibility and block installation when required cluster dependencies are missing or incompatible. 

Third, skill package managers can provide an audit command analogous to \texttt{npm audit}~\cite{npmDocs}. A \texttt{skill audit} can resolve the full skill dependency graph and report confirmed vulnerable dependencies, including affected versions and evidence channels.

\textbf{For skill developers.} Developers can maintain a lockfile-like record during skill development. Community discussions have already shown that, when a reused skill dependency does not preserve its source correctly, downstream users can lose the context needed to reinstall or audit it~\cite{removeDuplicateFrontendDesignSkill,experimentalInstallSubpathBug}. A skill lockfile can therefore be maintained whenever a skill introduces another dependency. It can record exact dependency versions, source repositories, paths, and optional status, so that downstream users or package managers can reproduce the dependency graph and audit inherited risks.

\subsection{Threats to Validity}

\textbf{Construct validity.} 
\tool{} records only confirmed dependency edges as dependencies, retaining low-confidence matches as annotations that are excluded from analysis. Following conventional software supply chain practice~\cite{latendresse2022not,alia2026you}, we consider dependencies beyond runtime requirements, including optional and other declared dependency types, and therefore include skill, package, and service-use dependencies in our analysis. The security-relevant patterns in RQ3 are audit signals rather than confirmed vulnerabilities, with key flagged cases manually inspected.

\textbf{Internal validity.} 
\tool{} is both the artifact and the measurement instrument for RQ2 and RQ3, so inaccuracies in dependency extraction may affect the reported findings. We mitigate this threat by evaluating \tool{} on the independently annotated SKILL-DEP benchmark, retaining supporting evidence for every extracted dependency, and manually inspecting representative flagged cases.

\textbf{External validity.} 
Our study covers GitHub-backed SkillsMP skill snapshots, excluding private and enterprise skills. Although SKILL-DEP is constructed using stratified sampling, its size may not fully represent the diversity of the whole skill corpus, particularly long-tail domains and uncommon dependency patterns. In addition, the rapid growth of skills means structural patterns may shift.

\section{Conclusion}
\label{sec:conclusion}

Agent skills are becoming reusable software artifacts, but their dependencies remain largely implicit. This paper introduces Agent Skill Supply Chains (ASSCs) as mixed skill-package-service dependency graphs and presents \tool{}, which reconstructs ASSCs from skills and emits SkillBOM. Evaluation on the SKILL-DEP benchmark shows that \tool{} recovers ASSCs accurately and comprehensively. 
Applying \tool{} to 1.43 million skills, we characterize the structure of ASSCs and their security patterns. We find four structural patterns: ASSCs are poorly governed despite being activation-ready; ASSCs span skill, package, and service dependencies with concentrated reuse; recursive skill reuse creates dependency expansion and hidden package inventory; and skill dependency clusters are formed around related workflows. 
We also find that inspecting only the root skill is insufficient, because security-relevant skills, packages, and services can reach downstream skills even when those downstream skills never mention them directly.
These findings call for graph-aware ASSC governance: typed dependency manifests, dependency-cluster management, risk-warning audit commands for skill infrastructure maintainers, and lockfile-like records for skill developers.

\bibliographystyle{IEEEtran}
\bibliography{references}

\end{document}